\documentstyle[amssymb,secnumtab]{fbssuppl}
\title{Questions on the Quark Model: Panel Discussion
\thanks{Presented by NCM at the Trento N$^*$ workshop}}
\author{Nimai C. Mukhopadhyay\thanks{{\it E-mail
address : mukhon@rpi.edu}} and R.M. Davidson \thanks{{\it E-mail
address : davidr@rpi.edu}}}
\institute{Department of Physics, Applied Physics and Astronomy,
Rensselaer Polytechnic Institute,
Troy, New York 12180-3590}

\begin{document}

\maketitle

In atomic, nuclear, and particle physics, one often deals with
complicated potentials where the analytic solution to the bound
state wave functions is not known. One general approach to this
problem is to diagonalize the Hamiltonian in a finite, i.e., truncated,
model space. While this method is generally very good in determining the
low-lying energy levels, it is not without problems. For example,
an operator that is known to commute with the Hamiltonian may not,
in the model space, actually commute with the Hamiltonian. To deal
with this problem, one tries to find effective operators in the model
space that do satisfy the properties of the operators in the untruncated
space.

A specific example of this problem is current conservation in the
quark model, which was discussed at length at this workshop \cite{viv,alf}.
One problem is that due to truncation of the space, the current is not
conserved, i.e., $\partial_{\mu}j^{\mu}\ne 0$. Of course, one can always
define a new current that is conserved, e.g., in momentum space,
\begin{equation}
\tilde{j}^{\mu} = j^{\mu} - { K\cdot j \over K^2}K^{\mu} \; ,
\end{equation}
where $j$ is the current and $K$ is the photon four-momentum.
However, this is not unique, which is not satisfactory since we want
a microscopic model to give us an unique answer.

The degree of violation of current conservation in the quark model
has been investigated by Bourdeau and Mukhopadhyay \cite{bor} and
Drechsel and Giannini \cite{dre}. The general conclusion from these
works is that there is a severe violation of current conservation,
making predictions of small quantities, such as the $N-\Delta$(1232)
$E2$ transition, unreliable. Indeed, Buchmann \cite{alf} presented evidence
at this workshop that two-body currents, neglected in most quark model
calculations of $E2$, actually dominate this transition amplitude.

Close and Li \cite{li} have proposed general effective operators to deal with
this problem. Capstick and collaborators \cite{cap} have partially
implemented these operators in their relativised quark model, but
even Capstick admitted at this workshop that $E2$ calculations are
still not very accuarte. The problem in these calculations is not
just the truncation of the space. In fact, compared to the
earliest quark model calculations, the model spaces now being used are
huge. The additional problem one is now encountering is what to use
for the electromagnetic transition operators. In other words, how do
we {\it consistently} relativize these operators? In atomic physics, this
may be done by means of a Foldy-Wouthuysen (FW) transformation.
Unfortunately, in quark model calculations the FW series converges
very slowly, if at all \cite{zha}. A specific example of this may
be found in the classic text book by Close \cite{close},
where, in the FW scheme,
the magnetic interaction energy turns out to be nearly as large as
the Coulomb interaction energy.

In conclusion, this is a difficult problem that deserves serious
attention. the good news is that it is being taken seriously as
was evidenced by the presentations of Demetriou \cite{viv} and
Buchmann \cite{alf} at this workshop.

We thank the participants of this panel for many stimulating
discussions.

\bibliographystyle{unsrt}

\end{document}